\documentclass[twocolumn,a4paper,showpacs,preprintnumbers,amsmath,amssymb,prl]{revtex4}

\usepackage{graphicx} 
\usepackage{dcolumn}
\usepackage{bm} 
\usepackage{xcolor}

\begin{document}

\hyphenation{te-tra-go-nal}

\bibliographystyle{apsrev}

\title{Optical conductivity study of screening of many-body effects in graphene interfaces}

\author{Pranjal Kumar Gogoi$^{1,2,3}$, Iman Santoso$^{1,3,4}$, Surajit Saha$^{1,2}$, Sihao Wang$^{2}$, Antonio H. Castro Neto$^{2,4}$,  Kian Ping Loh$^{1,4,5}$, T. Venkatesan$^{1,2,6}$, Andrivo Rusydi$^{1,2,3}$}

\affiliation{$^{1}$NUSNNI-NanoCore, National University of Singapore, Singapore 117576}
\affiliation{$^{2}$Department of Physics, National University of Singapore, Singapore 117542}
\affiliation{$^{3}$Singapore Synchrotron Light Source, National University of Singapore, 5 Research Link, Singapore 117603, Singapore}
\affiliation{$^{4}$Graphene Research Centre, Faculty of Science, National University of Singapore, Singapore 117546}
\affiliation{$^{5}$Department of Chemistry, National University of Singapore, Singapore 117543}
\affiliation{$^{6}$Department of Electrical and Computer Engineering, National University of Singapore, Singapore 117576}
\date{\today}

\begin{abstract}

Theoretical studies have shown that electron-electron (e-e) and electron-hole (e-h) interactions play important roles in many observed quantum properties of graphene making this an ideal system to study many body effects. In this report we show that spectroscopic ellipsometry can enable us to measure this interactions quantitatively. We present spectroscopic data in two extreme systems of graphene on quartz (GOQ), an insulator, and graphene on copper (GOC), a metal which show that for GOQ, both e-e and e-h interactions dominate while for GOC e-h interactions are screened. The data further enables the estimation of the strength of the many body interaction through the effective fine structure constant, $\alpha_{g}^{*}$. The $\alpha_{g}^{*}$ for GOQ indicates a strong correlation with an almost energy independent value of about 1.37. In contrast, $\alpha_{g}^{*}$ value of GOC is photon energy dependent, is almost two orders of magnitude lower at low energies indicating very weak correlation.

\end{abstract}


\maketitle

Tailoring the dielectric interface has been proposed as an immediate step to manipulate the electron-electron (e-e) and electron-hole (e-h) interactions in graphene \cite{AntonioRMP} as it is two dimensional in structure and also in almost all cases requires a substrate \cite{MakPRL, ChaeNanL, FeldmanNatPhys, WeitzSci, BolotinNat, ZhangNat}. Experimental reports addressing questions whether graphene is a strongly or weakly interacting system have thrown light on the fact that the answer mostly depends on the energy scale of interest \cite{ReedScien, BostwickScien}. Optical measurements from far infrared to deep ultra-violet (0.1- 5.5 eV) have been performed on graphene on insulating substrates \cite{MakPRL,KravetsPRB}(with negligible screening effect) and on free-standing grapheme \cite{ChaeNanL} which show prominent contributions from both e-e and e-h interactions as predicted by theoretical calculations\cite{YangPRL}. An extreme case of screening by a substrate dielectric interface is to use a metallic substrate, which has a huge supply of free electrons, in close proximity with the graphene layer. For example, metallic substrates have been used for one dimensional system like single walled nanotubes (SWNT) to study the screening of many body effects\cite{LinNatMat}. Optical conductivity measurement is not only sensitive to interband transitions and intraband (e.g Drude, e-e interactions) processes but it is also the most direct way to observe the effect of e-h interactions as in optical processes electron excitation creates a concomitant hole state.

The effects of different dielectric environments on the transport properties of graphene highlight the importance of the effective fine structure constant ($\alpha_{g}^{*}$) as one of the parameters of screening \cite{JangPRL}. The fine structure constant ($\alpha_{g}$) is the ratio of  potential energy to kinetic energy of electrons for free standing graphene, and is $\frac{e^{2}}{\hbar V_{f}}$ , where $\hbar$ is Planck’s constant and $\hbar V_{f}$ is the renormalized Fermi velocity near Dirac point, has the nominal value of 2.2  and it  indicates that graphene is a strongly interacting system \cite{AntonioRMP,WehlingPRL}. However a recent study on highly oriented pyrolytic graphite\cite{ReedScien} has elucidated the fact that the effective fine structure constant ($\alpha_{g}^{*}$) which is given by  $\alpha_{g}^{*}(k,\omega)=\frac {\alpha_{g}}{\varepsilon(k,\omega)}$, where k and  $\omega$  are momentum and energy respectively, may deviate from the value of 2.2 to far lower magnitude – indicating that graphene might be weakly interacting depending on k and $\omega$. On the other hand when the graphene layer is sandwiched between two dielectric mediums whose complex dielectric constants are  $\varepsilon^{1}$ and $\varepsilon^{2}$ ,  respectively the effective fine structure constant can be tuned and the new value is given by  $\alpha_{g}^{*}(k,ω)=\frac {2 e^{2}}{(\varepsilon^{1} + \varepsilon^{2})\hbar V_{f}}$.

In this letter, real part of the optical conductivity $\sigma_{1}(\omega)$ of monolayer  graphene  has been extracted from measured ellipsometric parameters $\Psi$ and $\Delta$.  The energy range of measurement is 0.5- 5.3 eV. The details of samples and measurement techniques are described in the Supporting Online Material \cite{supplementary}. Figure 1a shows that $\sigma_{1}(\omega)$ of the graphene layer on quartz (GOQ) (shown in red) has behaviour akin to exfoliated monolayer graphene in different regions of the energy range of measurement.  Beyond the infrared range ($\omega>$1.5 eV) $\sigma_{1}(\omega)$ starts gradually increasing from the constant value  $\frac {\pi e^{2}} {2h}$.  It may be mentioned that our observed constant value is consistent with other CVD grown graphene\cite{LeeAPL} which is slightly less than the   universal value observed in case of pristine exfoliated graphene\cite{MakPRL, ChaeNanL, NairSciec}.

A prominent asymmetric peak in $\sigma_{1}(\omega)$ is observed at  4.6 eV which can be attributed to excitonic effects in the optical  transitions at the M point in the Brillouin zone of  graphene. This peak is a result of the interplay between interband transitions, e-e and e-h interactions\cite{YangPRL}.  If one considers only band to band transitions using local density approximation (LDA) approach, the optical transition peak should  occur at ~4.1 eV. By inclusion of the e-e interactions through GW approach, the optical transition peak is predicted at 5.2 eV. By further incorporating the e-h interaction the optical transition peak is predicted to be red shifted by ~600 meV from 5.2 eV to 4.6 eV\cite{YangPRL}.

\begin{figure}
\begin{center}
\includegraphics[width=3.4in]{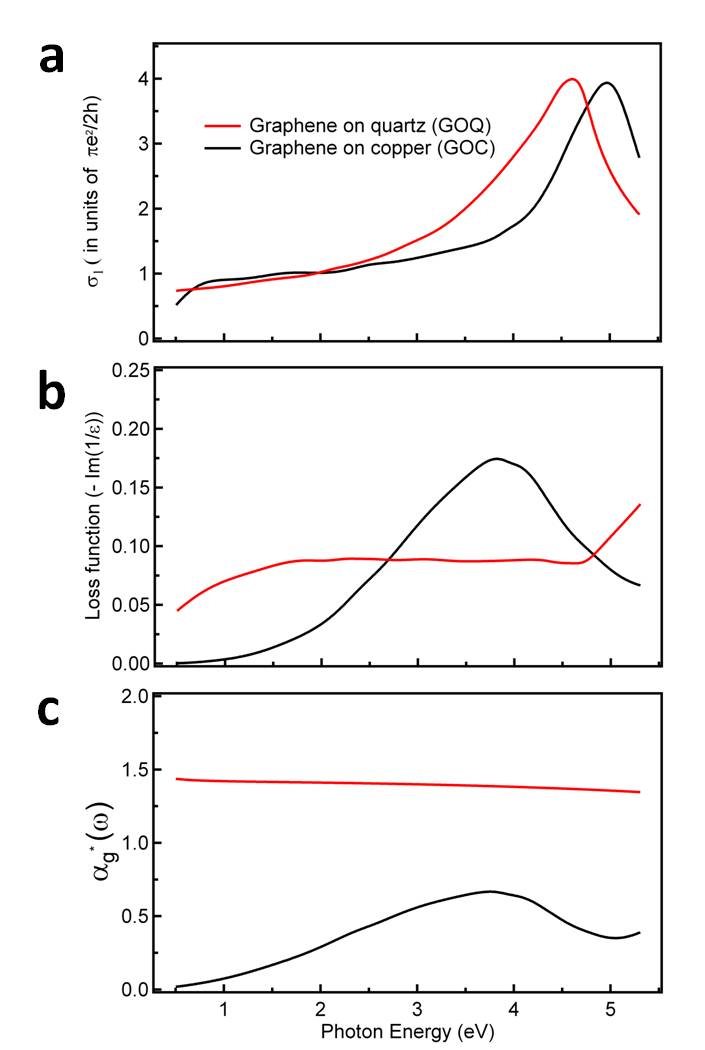}
\caption{\label{fig:fig1-R} (a) Real part of optical conductivity, $\sigma_{1}(\omega)$ of graphene on quartz (graphene on copper) shown in red (black). (b) Loss function of graphene on quartz (graphene on copper) shown in red (black). (c) Energy dependence of effective fine structure constant, $\alpha_{g}^{*}$, of graphene on quartz (graphene on copper) shown in red (black).}
\end{center}
\end{figure}

As shown in Figure~\ref{fig:fig1-R}(a) the most important observation of this study is that graphene on copper substrate (GOC, shown in black) has distinctively different trend of $\sigma_{1}(\omega)$ at the ultraviolet range. The peak for optical transitions for GOC is blue shifted to 4.96 eV compared to 4.60 eV as found in GOQ. Interestingly the line-shape of $\sigma_{1}(\omega)$ for GOC is  symmetric unlike the line shape of $\sigma_{1}(\omega)$ for GOQ which possesses asymmetric  profile. These two aspects $-$ red-shift of the optical transition peak and symmetric line-shape - are the key signatures of the different roles played by e-e and e-h interactions.

Figure~\ref{fig:fig1-R}(b) shows the Loss function ($- Im(\frac {1}{\varepsilon})$) for both GOQ and GOC. We do not see any structure for GOQ at 4.6 eV as seen in $\sigma_{1}(\omega)$. This is an evidence of the absence of plasmonic contribution to the peak in $\sigma_{1}(\omega)$ at 4.6 eV. For GOC we observe a broad structure centered at 3.9 eV which may be attributed to plasmonic excitations. However this is far below the peak seen for $\sigma_{1}(\omega)$ at 4.96 eV. This again rules out plasmonic contribution.

\begin{figure}
\begin{center}
\includegraphics[width=3.4in]{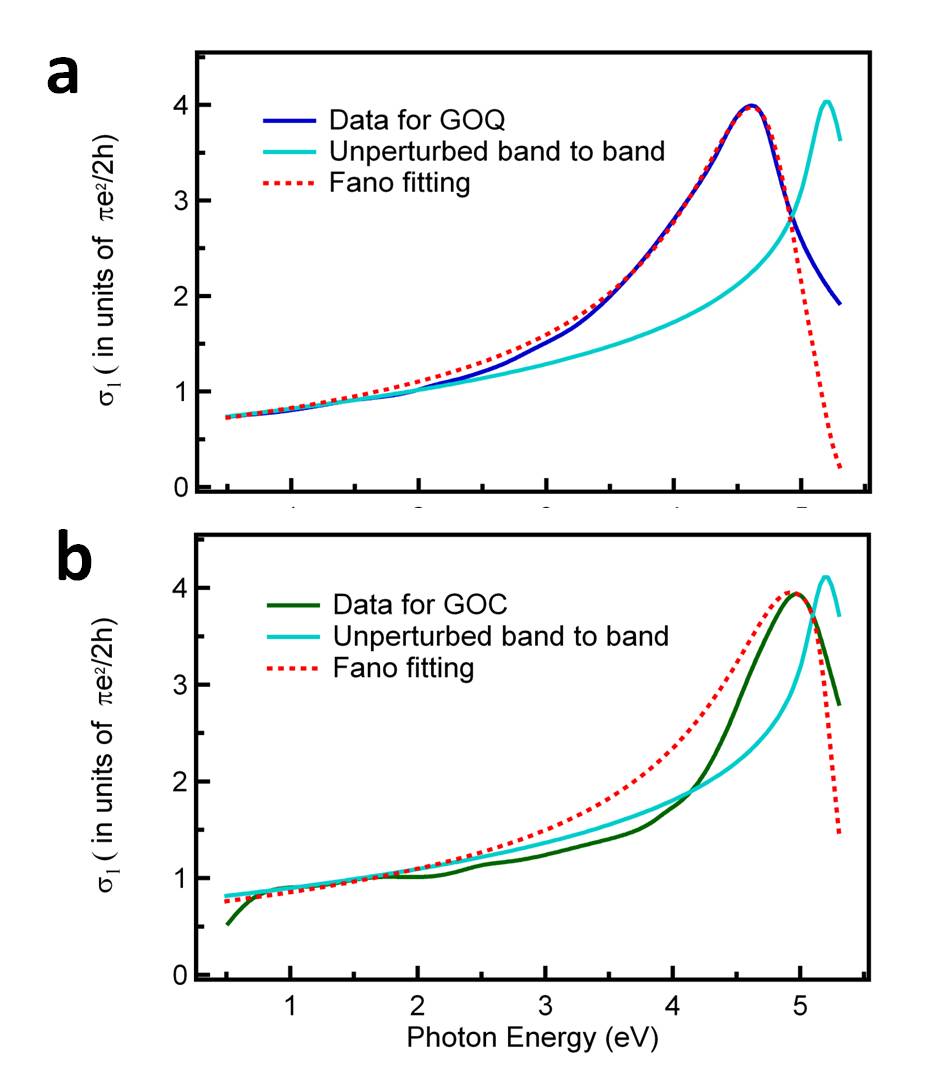}
\caption{\label{fig:fig2-S1-W} Fano line-shape analysis (a) for GOQ and (b) for GOC. Fit of GOQ  and GOC experimental $\sigma_{1}(\omega)$ are shown in red dashed lines. The $\sigma_{cont}(\omega)$ is taken as the unperturbed band to band transition shown in cyan. (See Supporting Online Material \cite{supplementary} for details of Fano line-shape analysis.) }
\end{center}
\end{figure}

This asymmetric line-shape of the $\sigma_{1}(\omega)$ can be interpreted using  a phenomenological approach proposed by Fano \cite{FanoPR, PhillipsPR, Yu, Phillips } which takes into account e-h interactions. In Fano theory discrete excitonic  states residing below an electronic continuum couple with the continuum states  giving rise to considerable asymmetry in the optical transition strengths near a saddle point singularity.  We fit our experimental data using Fano interference analysis by employing a phenomenological relationship where a dominant excitonic state is coupling with the continuum\cite{PhillipsPR,Yu}.  Figure~\ref{fig:fig2-S1-W} shows a detailed Fano analysis on the $\sigma_{1}(\omega)$ data for GOQ and GOC. The peak position and asymmetry of $\sigma_{1}(\omega)$  of GOQ can be well-fitted  (the slight  mismatch at 1.5 eV $-$ 3 eV may be due to inherent quality issue of CVD graphene and above 5 eV the match discrepancy can be attributed  to the use of unperturbed band to band transition result instead of exact GW result\cite{MakPRL} as the starting point)  using this Fano approach with the parameters q = -1.16, $\Gamma$ = 0.99 eV and  $E_{res}$ = 4.90 eV (see Supporting Online Material \cite{supplementary} for details).  On the contrary the peak position of $\sigma_{1}(\omega)$  for GOC is at 4.96 eV and with Fano parameters of q = -0.96, $\Gamma$ = 0.98 eV and  $E_{res}$ = 5.19 eV we can account for the redshift from the unperturbed peak at 5.2 eV but the distinctively symmetric shape of our result cannot be fitted with this model as seen from  Figure~\ref{fig:fig2-S1-W}(b). Therefore it signifies the very weak strength, if any, of excitonic contribution in this redshift\cite {Phillips}. This further indicates strong screening of e-h interactions in GOC.

More generally screening effects can be quantified by analyzing the $\alpha_{g}^{*}$. In Figure~\ref{fig:fig1-R}(c) we show  $\alpha_{g}^{*}(\omega)$ for GOQ and GOC.  Here we find that for GOQ, $\alpha_{g}^{*}(\omega)$  has a value of  1.37  at 4.6 eV and it does not vary by more than $5$ in the energy range of our interest (0.5- 5.3 eV). This dynamic $\alpha_{g}^{*}(\omega)$ is greater than the static  $\alpha_{g}^{*}(\omega=0)$= 0.81 which is regularly used to describe correlations in graphene (on quartz or SiO2/Si substrate). This indicates that the dynamical screening in GOQ is weak and so the system can be categorized as a strongly interacting system for this energy range. In the case of GOC, $\alpha_{g}^{*}$ at 4.96 eV is ~0.36. It is interesting to note that in contrast to GOQ,  $\alpha_{g}^{*}$  varies from 0.02 to 0.67 in the energy range of 0.5 – 5.3 eV for GOC. This basically means that the dynamic screening is stronger than in the case of GOQ as indicated by the lower value of $\alpha_{g}^{*}(\omega)$. These results suggest that GOC is a weakly interacting system in comparison to GOQ.

\begin{figure}
\begin{center}
\includegraphics[width=3.4in]{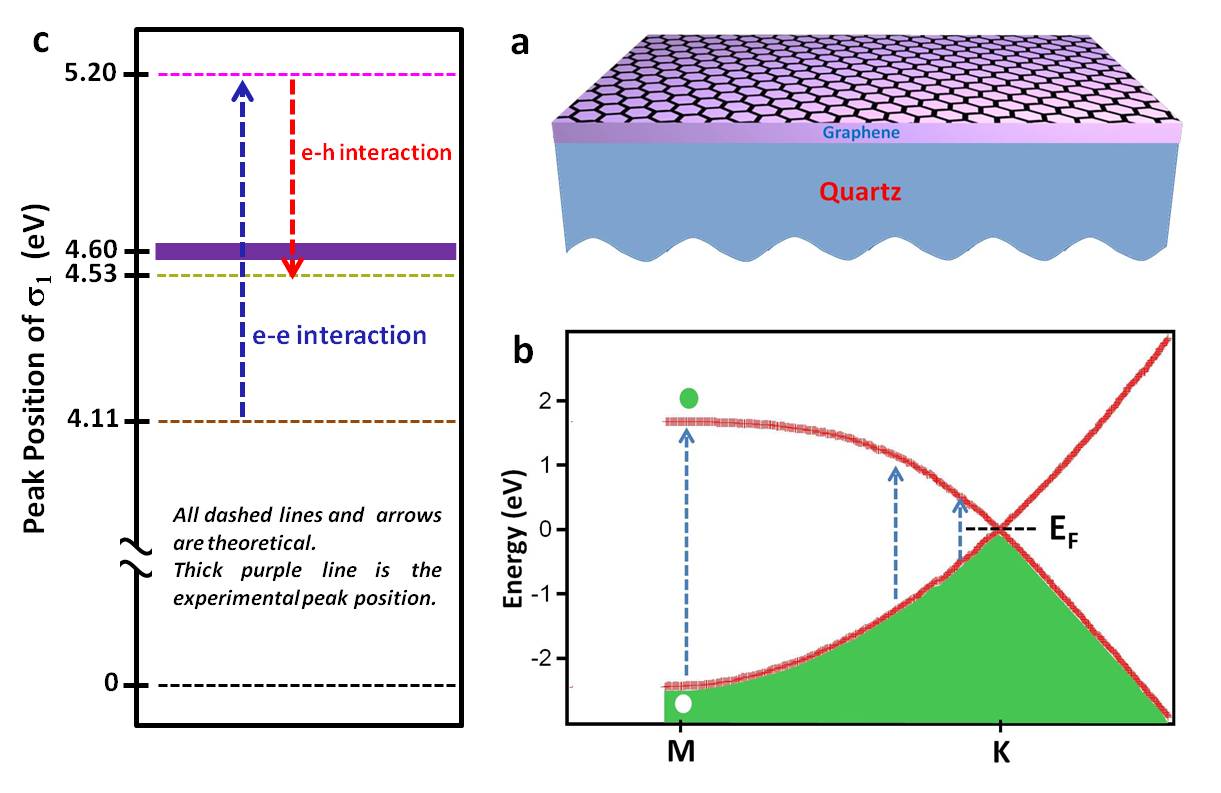}
\caption{\label{fig:fig3-model} (a) Cartoon of graphene on quartz substrate (GOQ). (b) Schematic band diagram from Density functional theory (DFT) of graphene on quartz (GOQ) assuming no doping as in free-standing case. Possible optical transitions are shown where all states above the Dirac point are accessible. (c) Schematic energy values that represent the peak position in the optical transitions for GOQ. All dashed lines and arrows are representing theoretical predictions. The brown dashed line(\color {brown}{---}\color {black}) at 4.11 eV represents the result from local density approximation (LDA) theory\cite{YangNanL} which accounts only for band to band transitions. The pink dashed line (\color {pink}{---}\color {black}) at 5.20 eV represents the result from GW calculations which takes into account many-body electron –electron (e-e) interactions\cite{YangNanL}. The blue arrow (\color {blue}{$\rightarrow$}\color {black}) indicates the difference in energy of GW from LDA calculations. The green dashed line (\color {green}{---}\color {black}) at 4.53 eV represents the energy value predicted by GW- Bethe Salpeter Equation  (GW-BSE) approach which includes also the electron –hole (e-h) interactions\cite{YangNanL}. The red arrow (\color {red}{$\rightarrow$}\color {black} ) indicates the energy difference between GW and GWBSE calculations which is the contribution only from e-h interactions. The thick purple line (\color {purple}{---}\color {black}) at 4.60 eV is representing the asymmetric peak position in  our experimental result described in text for GOQ. The thickness is proportional to the error bar which is affected mostly by the fitting procedure.The closeness of our result with the GW-BSE prediction indicates the presence of  both e-e and e-h interactions in GOQ.}
\end{center}
\end{figure}

\begin{figure}
\begin{center}
\includegraphics[width=3.4in]{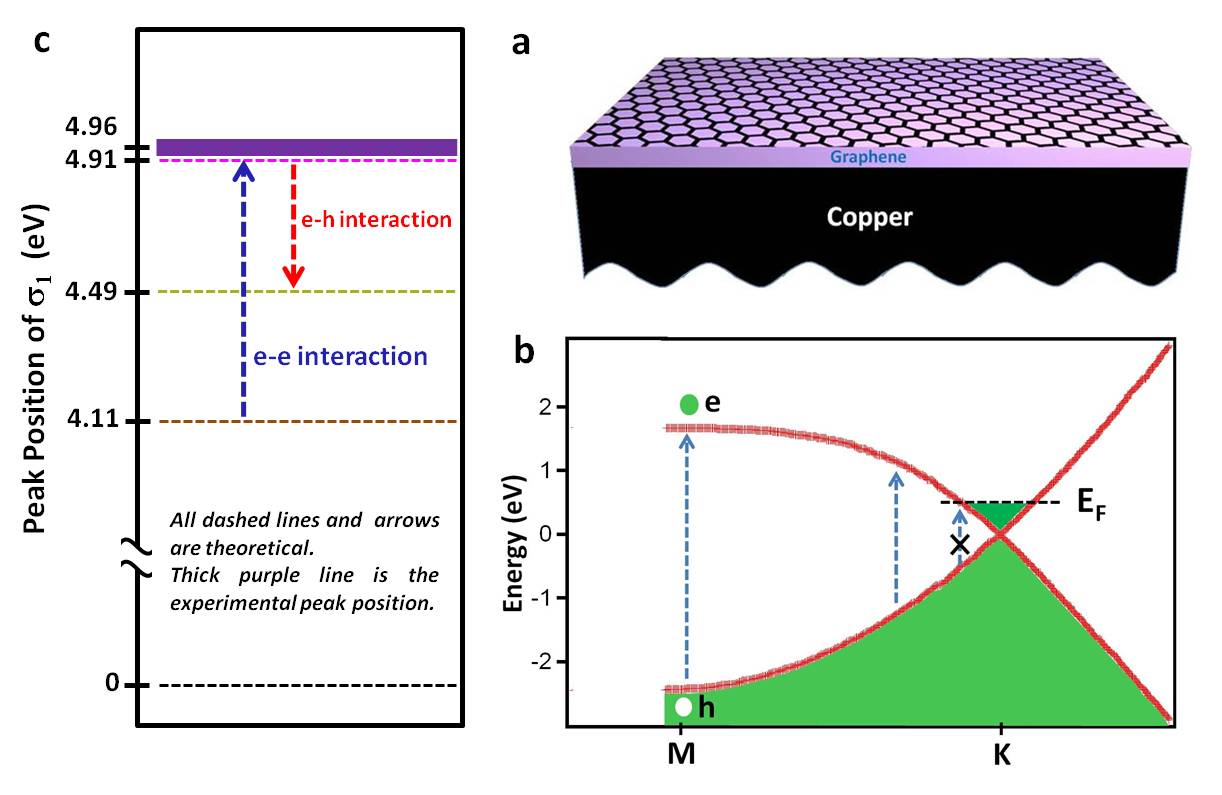}
\caption{\label{fig:fig4-model} (a) Cartoon of graphene on copper substrate (GOC). (b) Schematic band diagram from Density functional theory (DFT) of graphene on copper (GOC) assuming considerable electron doping based on DFT calculations\cite{GiovannettiPRL}. Possible transitions are shown with cyan arrows ( \color {cyan}{$\rightarrow$}\color {black} ) to states above the Fermi level. States below the Fermi level but above the Dirac point are not accessible (represented by crossed cyan arrow). (c) Schematic energy values that represent the peak positions in the optical transitions for GOC. All dashed lines and arrows are representing theoretical predictions where electron doped graphene (0.01 electrons per unit cell of graphene) is considered.  The brown dashed line (\color {brown}{---}\color {black}) at 4.11 eV represents the result from local density approximation (LDA) theory which accounts only for band to band transitions\cite{YangNanL}. The pink dashed line (\color {pink}{---}\color {black}) at 4.91eV represents the result from GW calculations which takes into account many-body e-e interactions\cite{YangNanL}. The blue arrow ( \color {blue}{$\rightarrow$}\color {black} ) indicates the difference in energy of GW from LDA calculations. The green dashed line  (\color {green}{---}\color {black}) at 4.49 eV represents the energy value predicted by GW- Bethe Salpeter Equation  (GW-BSE) approach which includes also the e-h interactions\cite{YangNanL}. The red arrow (\color {red}{$\rightarrow$}\color {black} ) indicates the energy difference between GW and GWBSE calculations which is the contribution only from e-h interactions. The thick purple line (\color {purple}{---}\color {black}) at 4.96 eV is representing the symmetric peak position in $\sigma_{1}(\omega)$ of our experimental result described in text for GOC. The thickness is proportional to the error bar which is affected mostly by fitting procedure. The fact that our experimental peak position is close to GW prediction indicates that e-h interactions are screened almost fully}
\end{center}
\end{figure}

In Figures~\ref{fig:fig3-model} and ~\ref{fig:fig4-model} we present  schematic energy diagrams and optical transition levels in order to explain the observed optical conductivity and possible scenarios for both GOQ and GOC respectively.  For GOQ, $\sigma_{1}(\omega)$ peak is at  4.6 eV  and it  can be explained with the existence of both e-e and e-h interactions which are well supported by theoretical model\cite{YangPRL}. This suggests that the interactions between graphene and the substrate are weak and the graphene layer behaves almost like free-standing graphene.

For GOC, $\sigma_{1}(\omega)$  is at 4.96 eV and this 360 meV blue shift (from the GOQ peak) is considerably large. This may involve two processes $-$ firstly, considerable electron transfer from the metal substrate\cite{GiovannettiPRL} giving rise to electron doping and secondly, the effect of this electron doping on the optical conductivity\cite{YangNanL}.  Theoretically, for graphene deposited on metal (like Cu, Ni)\cite{GiovannettiPRL}  the Fermi level  shifts considerably depending on the  work function of the metal as well as the separation between the metal substrate and the graphene layer. For our case, graphene is grown on copper without any other layer in between (see Supporting Online Material\cite{supplementary}) with a gap of $~$ 0.3 nm. This electron doping screens the e-h interactions while e-e interactions are dominant. The scenario depicted in Figure~\ref{fig:fig4-model} is from predictions using DFT calculations of a Fermi level shift of 0.5 eV above the Dirac point when graphene is in contact with copper substrate\cite{GiovannettiPRL}. For the case of free standing doped graphene one would expect to see the excitonic peak at almost similar position like in GOQ at 4.49 eV\cite{YangNanL} due to the presence of both e-e and e-h interactions even while doped.

Another noticeable aspect of our result is the absence of distinct signature of the coupling between the $p_{z}$ orbital of the carbon atoms and $d_{z2}$ orbital of the copper atoms in  $\sigma_{1}(\omega)$. Theoretical studies\cite{GiovannettiPRL,ZhipingJP} have predicted that metals can be broadly divided into two classes depending on the coupling strength of these orbitals. For example in case of graphene on nickel, graphene bandstructure is perturbed heavily due to strong coupling, whereas in case of graphene on copper (which belongs to the other class of metals) graphene still retains most of its intrinsic bandstructure features. However our result shows even smaller coupling in case of graphene on copper than predicted. Firstly we do not see any structure below 5 eV for the $\sigma_{1}(\omega)$  of GOC. This is in  contrast to theoretical calculations which predicted  presence of  copper $d_{z^2}$ bands at approximately 2 eV below the Fermi energy when the substrate is Cu (111)\cite{ZhipingJP}. Secondly we do not see a Pauli blockade in   $\sigma_{1}(\omega)$  for GOC at about 1 eV which translates into about 0.5 eV of Fermi level shift due to charge transfer. Rather we see a dip in the $\sigma_{1}(\omega)$ plot at about 0.5 eV which represents a Fermi level shift of less than 0.25 eV.  These two aspects may be a signature of weak coupling between  polycrystalline  copper and graphene.

In conclusion, we observe that the dynamical screening in graphene on metallic substrate is stronger  than in the case of graphene on quartz by as large as two orders of magnitude.  We propose that the observed blue-shift in the peak position of optical conductivity at the van Hove singularity (at the M point) is the result of the fact that  electron-electron  interactions are still dominating  but    electron-hole interactions are strongly screened. Our result opens  new paths to study the interplay of e-e and e-h interactions and their individual strengths in many-body physics. Furthermore it demonstrates the suitability of spectroscopic ellipsometry technique to reveal the interactions in graphene interfaces.

\acknowledgments
We would like to acknowledge the discussion with Vitor Manuel Pereira.  This work is supported by NRF-CRP grant Tailoring Oxide Electronics by Atomic Control, MOE Tier 2, NUS YIA, NUS cross faculty grant, FRC, Advance Material (NanoCore) R-263-000-432-646.


\begin{thebibliography}{}

\bibitem{AntonioRMP} A. H. Castro Neto {\it et.al}, Rev. Mod. Phys {\bf 81}, 109 (2009).
\bibitem{MakPRL} K. F. Mak {\it et.al}, Phys. Rev. Lett. {\bf 106}, 046401 (2011).
\bibitem{ChaeNanL} D. H. Chae {\it et.al}, Nano. Lett. {\bf 11}, 1379 (2011).
\bibitem{FeldmanNatPhys}  B. E Feldman {\it et.al}, Nature Phys. {\bf 5}, 889 (2009).
\bibitem{WeitzSci} R. T. Weitz {\it et.al}, Science {\bf 330}, 812 (2010).
\bibitem{BolotinNat} K. I. Bolotin {\it et.al}, Nature {\bf 462}, 196 (2009).
\bibitem{ZhangNat} Y. Zhang {\it et.al}, Nature {\bf 438}, 201 (2005).
\bibitem{ReedScien}	J. P. Reed {\it et.al}, Science {\bf 330}, 805 (2010).
\bibitem{BostwickScien} A. Bostwick {\it et.al}, Science {\bf 328}, 999 (2010).
\bibitem{JangPRL} C. Jang {\it et.al}, Phys. Rev. Lett. {\bf 101}, 146805 (2008).
\bibitem{KravetsPRB} V. G. Kravets {\it et.al}, Phys. Rev. B. {\bf 81}, 155413 (2010).
\bibitem{YangPRL} L. Yang {\it et.al}, Phys. Rev. Lett. {\bf 103}, 186802 (2009).
\bibitem{LinNatMat} H. Lin {\it et.al}, Nature Mat. {\bf 9}, 235 (2010).
\bibitem{LeeAPL} C. Lee {\it et.al}, Appl. Phys. Lett. {\bf 98}, 071905 (2011).
\bibitem{supplementary} See Supporting Online Material.
\bibitem{NairSciec} {R. R. Nair} {\it et.al}, Science  {\bf 320},  1308 (2008).
\bibitem{WehlingPRL} T. O. Wehling {\it et.al}, Phys. Rev. Lett. {\bf 101}, 026803 (2008).
\bibitem{GiovannettiPRL}C. Giovannetti {\it et.al}, Phys. Rev. Lett. {\bf 101}, 026803 (2008).
\bibitem{YangNanL} L. Yang, Nano. Lett. {\bf 11}, 3844 (2011).
\bibitem{FanoPR} U. Fano, Phys. Rev. {\bf 124}, 1866 (1961).
\bibitem{PhillipsPR} J. C. Phillips, Phys. Rev. {\bf 136}, A1705 (1964).
\bibitem{Yu} P.Y. Yu and  M. Cardona, Fundamentals of Semiconductors: Physics and Materials Properties  (Springer, Berlin, 1996).
\bibitem{Phillips} J. C. Phillips, Excitons. In The Optical Properties of Solids,  Editor: J. Tauc (Academic Press, New York, 1966).
\bibitem{ZhipingJP}X. Zhiping {\it et.al}, J. Phys.: Condens. Matter {\bf 22}, 485301 (2010).


\end{thebibliography}
\end{document}